\documentstyle{article}
\input amssymb.sty

\renewcommand{\thefootnote}{\fnsymbol{footnote}}
\def\H{{\cal H}}
\def\id{\mathop{\rm id}}
\def\F{{\cal F}}
\def\M{\real^{p+1}}

\def\real{{\Bbb R}}
\def\inprod#1#2{\left<#1,#2\right>}
\def\pder{\partial}

\def\ie{{\it i.e}.\ }
\def\eg{{\it e.g}.\ }
\def\inv#1{\frac{1}{#1}}

\begin{document}
\begin{titlepage}
\begin{center}
\noindent 26 March 2000\hfill DIAS-STP-00-03\\
\hfill {\tt hep-th/0003234}

\vspace{2cm}

{\large\bf Derivatives and the Role of the Drinfel'd Twist in
Noncommutative String Theory}

\vspace{1cm}

\setcounter{footnote}{0}
\renewcommand{\thefootnote}{\arabic{footnote}}

{\bf Paul Watts}\footnote{E-mail: {\tt watts@stp.dias.ie}, Tel: +353-1-614
0148, Fax: +353-1-668 0561}

\vspace{1cm}

Dublin Institute for Advanced Studies\\
School of Theoretical Physics\\
10 Burlington Road\\
Dublin 4\\
Ireland

\vspace{2cm}

{\bf Abstract}
\end{center}
\noindent We consider the derivatives which appear in the context of
noncommutative string theory.  First, we identify the correct derivations
to use when the underlying structure of the theory is a quasitriangular
Hopf algebra.  Then we show that this is a specific case of a more general
structure utilising the Drinfel'd twist.  We go on to present reasons as to
why we feel that the low-energy effective action, when written in terms of
the original commuting coordinates, should explicitly exhibit this
twisting.

\bigskip

\noindent PACS-99: 11.25.-w, 11.25.Hf\\
\noindent MSC-91: 46L87, 57T05, 81T30

\bigskip


\end{titlepage}
\newpage
\renewcommand{\thepage}{\arabic{page}}

\section{Introduction}
\setcounter{equation}{0}

Recent works on theories of open strings and D$p$-branes with a constant
nonvanishing Neveu-Schwarz 2-form $B_{ij}$ have suggested that the
noncommutativity which appears is an underlying and very general property
of such theories \cite{CL}--\cite{SW}.  Since Hopf algebras (HAs) often lie
at the root of noncommutative systems, we were motivated to look for a HA
structure for these theories, and showed that the noncommutative
$*$-product \cite{M}--\cite{F2} was in fact a specific case of a more
general multiplication defined in terms of the R-matrix $R$ of a
quasitriangular HA $\H$; furthermore, when $\H=\F$, where $\F$ was the HA
of functions on $\M$, as was the case for the aforementioned noncommutative
string theories, we found an explicit form for $R$ \cite{W} which covers
both the commutative ($B_{ij}=0$) and noncommutative ($B_{ij}\neq 0$)
cases.

However, it was not immediately apparent how we could introduce derivations
on the algebra endowed with this multiplication, $\widehat{\H}$.  One way
to think of derivations on a HA $\H$ is as elements of the dually paired HA
$\H^*$, with the action of the latter on the former given in terms of the
HA properties of both.  The problem was that $\widehat{\H}$ was shown {\em
not} to be a HA, and therefore neither was the dually paired space
$\widehat{\H}^*$.  This precluded the interpretation of the latter as local
derivatives on $\widehat{\H}$, so it was not immediately obvious how one
might define a gauge theory on the noncommutative space, since we needed a
derivative in order to construct the (noncommutative) field strength tensor
$\widehat{F}_{ ij}$ from the gauge field $\widehat{A}_i$, \ie $\widehat{F
}_{ij}=\pder_{[i}\widehat{A}_{j]}-i\widehat{A}_{[i}*\widehat{A}_{j]}$
\cite{SW}.  The question was, what could we use for $\pder_i$?  We
speculated that we might have to replace local derivatives by difference
operators, but this guess seemed to be contradicted by the fact that
regular derivatives were used consistently in \cite{SW}.

In this follow-up note to \cite{W}, we explain why the usual derivatives
are in fact the correct ones when dealing with noncommutative string
theory: We show that even though $\widehat{\H}$ is not a HA, there exists a
HA which has a well-defined action on $\widehat{\H}$ and plays the role of
the space of derivations.  This holds for arbitrary $\H$, and for the
specific case where $\H=\F$, this HA is $\F^*$ and the action is the same
as that for the usual partial derivative.  This is done in Section 2.

The ability to relate the commutative and noncommutative theories via the
R-matrix, however, turns out to be a bit of a fluke, being true only if
$\H^*$ is cocommutative.  While this is certainly true of the space of
derivations $\F^*$, if we want to be as general as possible, we must relax
this condition.  In Section 3, we demonstrate how this can be done by using
the Drinfel'd twist \cite{D1}, which allows us to find a generalisation of
the $*$-product and the space of derivations on the algebra constructed
with this $*$.  (Related but more mathematical treatments of this
construction may be found in \cite{GZ,BPvO}, and very recently \cite{O},
which covers much of the same in a broader context.)

However, using the Drinfel'd twist gives exactly the same derivatives and
noncommutative product as if we had used an R-matrix approach, so why pick
one over the other?  In Section 4, we present two arguments why we think
the former is more appropriate: First, the R-matrix construction connects
the commutative and noncommutative cases only when $\H$ is commutative (\ie
$\H^*$ is cocommutative), whereas the Drinfel'd twist includes both cases,
and therefore does not require us to make any {\it a priori} assumptions
about the algebraic structure of $\H$.  Secondly, an R-matrix must satisfy
two coproduct conditions, while the analogous element in the Drinfel'd
twisting only has to fulfill one, and there may be a way of naturally
implementing the latter using the Ward identity which must arise out of
gauge-fixing the form of $B_{ij}$.  These reasons lead us to think that the
Drinfel'd twist plays a fundamental role in noncommutative string theory,
specifically in helping to determine the form of the low-energy effective
action of the theory in terms of the {\em commutative} coordinates, \eg
Born-Infeld.

\bigskip

Throughout this letter we use terms and notations described in our previous
paper \cite{W}; the reader is referred therein for the details.

\section{The Leibniz Rule and the $*$-product}\label{R-leibniz}
\setcounter{equation}{0}

We begin with a HA $\H$ and its dually paired HA $\H^*$.  The (left) action
of $x\in\H^*$ on $f\in\H$ is given by $x\cdot f:=f_{(1)}\inprod{x}{
f_{(2)}}$, and satisfies the Leibniz rule $x\cdot(fg)=\left(x_{(1)}\cdot
f\right)\left(x_{(2)}\cdot g\right)$.  The elements of $\H^*$ (with the
exception of the unit 1 and its multiples) thus may be thought of as
derivations on $\H$.

In the case where $\H^*$ is quasitriangular with R-matrix $R$, we can
define a new multiplication between $f,g\in\H$ as
\begin{equation}
f*g:=f_{(1)}g_{(1)}\inprod{R_{21}}{f_{(2)}\otimes g_{(2)}}.\label{Rstar}
\end{equation}
$\widehat{\H}$ is taken to be the algebra equivalent to $\H$ as a vector
space and with the multiplication $*$.  It is {\em not} a HA, so neither is
the dually paired coalgebra $\widehat{\H}^*$.  We therefore cannot think of
$\widehat{\H}^*$ as derivations of $\widehat{\H}$.

However, let's go ahead and compute the action of $x\in\H^*$ on the product
$f*g$:
\begin{eqnarray}
x\cdot(f*g)&=&x\cdot\left(f_{(1)}g_{(1)}\right)\inprod{R_{21}}{f_{(2)}
\otimes g_{(2)}}\nonumber\\
&=&f_{(1)}g_{(1)}\inprod{x}{f_{(2)}g_{(2)}}\inprod{R_{21}}{f_{(3)}\otimes
g_{(3)}}\nonumber\\
&=&f_{(1)}g_{(1)}\inprod{\Delta(x)\otimes R_{21}}{f_{(2)}\otimes g_{(2)}
\otimes f_{(3)}\otimes g_{(3)}}\nonumber\\
&=&f_{(1)}g_{(1)}\inprod{\Delta(x)R_{21}}{f_{(2)}\otimes g_{(2)}}.
\end{eqnarray}
Using $\Delta(x)=R_{21}\left(\tau\circ\Delta(x)\right)R_{21}^{-1}$, we
obtain
\begin{eqnarray}
x\cdot(f*g)&=&f_{(1)}g_{(1)}\inprod{R_{21}\left(\tau\circ\Delta\right)(x)}{
f_{(2)}\otimes g_{(2)}}\nonumber\\
&=&f_{(1)}g_{(1)}\inprod{R_{21}\otimes\left(\tau\circ\Delta\right)(x)}{
f_{(2)}\otimes g_{(2)}\otimes f_{(3)}\otimes g_{(3)}}\nonumber\\
&=&f_{(1)}g_{(2)}\inprod{R_{21}}{f_{(2)}\otimes g_{(2)}}\inprod{x_{(2)}
\otimes x_{(1)}}{f_{(3)}\otimes g_{(3)}}\nonumber\\
&=&\left(f_{(1)}*g_{(2)}\right)\inprod{x_{(2)}}{f_{(2)}}\inprod{x_{(1)}}{
g_{(2)}}\nonumber\\
&=&\left(x_{(2)}\cdot f\right)*\left(x_{(1)}\cdot g\right).\label{leibniz}
\end{eqnarray}
So we see that the Leibniz rule is `reversed':  The first piece of the
coproduct of $x$ acts on the {\em second} function, and vice versa.

In \cite{W}, we reviewed the construction of the HA dually paired to $\H$,
denoted $\H^*$.  Recall that the coproduct and antipode for $x\in\H^*$ were
defined by
\begin{eqnarray}
\inprod{\Delta(x)}{f\otimes g}&:=&\inprod{x}{fg},\nonumber\\
\inprod{S(x)}{f}&:=&\inprod{x}{S(f)}.
\end{eqnarray}
However, this is not the only HA which may be constructed to the vector
space dual to $\H$: A different HA, called the {\it opposite dual} and
denoted $\H^{\rm op}$\footnote{This is the same well-known HA which plays a
key role in the construction of the Drinfel'd double \cite{D1}.}, can be
defined by keeping all the relations between $\H$ and $\H^*$ except the
above two, which are replaced by
\begin{eqnarray}
\inprod{\Delta'(x)}{f\otimes g}&:=&\inprod{x}{gf},\nonumber\\
\inprod{S'(x)}{f}&:=&\inprod{x}{S^{-1}(f)}.
\end{eqnarray}
We see that the coproduct on $\H^{\rm op}$ is the one on $\H^*$ with the
two spaces flipped, \ie $\Delta'=\tau\circ\Delta$.  From (\ref{leibniz}) we
can see that although $\H^*$ does not have a action on $\widehat{\H}$,
$\H^{\rm op}$ does, because
\begin{equation}
x\cdot(f*g)=\left(x_{(1)'}\cdot f\right)*\left(x_{(2)'}\cdot
g\right),
\end{equation}
where we have used the notation $\Delta'(x):=x_{(1)'}\otimes x_{(2)'}$.
Hence, we conclude that $\H^{\rm op}$, not $\H^*$ or $\widehat{\H}^*$, is
the space of derivations on $\widehat{\H}$.  And since in general $\H$ is
not cocommutative, the Leibniz rule on $\widehat{\H}$ does not have the
same form as that on $\H$.  However, if $\Delta'=\Delta$, \ie $\H^*$ is
cocommutative, then $\H^{\rm op}$ and $\H^*$ {\em are} the same, and the
space of derivations is the same for $\widehat{\H}$ as for $\H$.

\bigskip

If we now look at noncommutative string theory, the algebra $\H$ is the
function algebra $\F$ spanned by monomials in the coordinate maps $x^i$
taking the D$p$-brane into $\real^{p+1}$.  The $*$-product is introduced by
using the R-matrix
\begin{equation}
R:=e^{-\frac{i}{2}\theta^{ij}\pder_i\otimes\pder_j},\label{R-matrix}
\end{equation}
where $\theta^{ij}$ is related to $B_{ij}$ and the open string metric
$g_{ij}$ by \cite{SW}
\begin{equation}
\theta^{ij}:=-\left(2\pi\alpha'\right)^2\left(\inv{g+2\pi\alpha'B}B\inv{g
-2\pi\alpha'B}\right)^{ij}.
\end{equation}
This $R$ gives the (noncommutative) product between functions $f$ and $g$
as
\begin{equation}
f(x)*g(x)=\left.e^{\frac{i}{2}\theta^{ij}\frac{\pder^2}{\pder\xi^i\pder
\zeta^j}}f(x+\xi)g(x+\zeta)\right|_{\xi=\zeta=0}.\label{star}
\end{equation}
The dually paired HA $\F^*$ is the space spanned by monomials of the
partial derivatives $\pder_i$.  The coproduct is generated by $\Delta
\left(\pder_i\right)=\pder_i\otimes 1+1\otimes\pder_i$, and with the action
on $\F$ being the usual derivative, the Leibniz rule is the familiar
\begin{equation}
\pder_i\left(f(x)g(x)\right)=\left(\pder_if(x)\right)g(x)+f(x)\left(\pder_i
g(x)\right)
\end{equation}
(the $\cdot$ signifying the action has been suppressed in the above two
equations).  But since this HA is cocommutative, this is also the Leibniz
rule for the action of $\F^{\rm op}$ on $\F$.  Hence, $\F^*=\F^{\rm op}$,
and this is the reason that one can use the familar derivatives even when
the space is noncommutative, as was done in \cite{SW}.

\section{The Drinfel'd Twist}
\setcounter{equation}{0}

The material in the preceding Section is in fact a specific example of a
more general construction: Suppose $\H$ is a HA such that there exists an
invertible element $F\in\H\otimes\H$ which satisfies $(\epsilon\otimes
\id)(F)=(\id\otimes\epsilon)(F)=1$ as well as the coproduct identity
\begin{equation}
F_{12}\left(\Delta\otimes\id\right)(F)=F_{23}\left(\id\otimes\Delta\right)
(F).\label{FYBE}
\end{equation}
If this is the case, then a new HA $\H^F$, called the {\it Drinfel'd twist}
of $\H$ \cite{D1}, can be defined in the following way: $\H^F=\H$ at the
algebra level, and the counit and unit of $\H^F$ are the same as those of
$\H$.  The coproduct and antipode, however, are given in terms of those on
$\H$ by
\begin{eqnarray}
\Delta^F(f):=F\Delta(f)F^{-1},&&S^F(f):=\sigma^{-1}S(f)\sigma,
\end{eqnarray}
where $\sigma$ is the quantity constructed from $F:=F_{\alpha}\otimes
F^{\alpha}$ (sum implied) via
\begin{equation}
\sigma:=m\left((\id\otimes S)\right)(F)\equiv F_{\alpha}S\left(F^{\alpha}
\right).
\end{equation}
(The inverse can be shown to be $\sigma^{-1}=m\left((S\otimes\id)\left(
F^{-1}\right)\right)$.)  For future reference, we also use the notation
$\Delta^F(f):=f_{(1)F}\otimes f_{(2)F}$.

Now suppose we start with dually paired HAs $\H$ and $\H^*$, and an element
$F$ in $\H^*\otimes\H^*$ satisfying (\ref{FYBE}) exists; then a new
product, $*$, may be defined on $\H$ via
\begin{equation}
f*g:=f_{(1)}g_{(1)}\inprod{F^{-1}}{f_{(2)}\otimes g_{(2)}}.\label{Fstar}
\end{equation}
We can then check associativity by first computing the triple product
$(f*g)*h$:
\begin{eqnarray}
(f*g)*h&=&\left(f_{(1)}g_{(1)}\right)*h\inprod{F^{-1}}{f_{(2)}\otimes
g_{(2)}}\nonumber\\
&=&f_{(1)}g_{(1)}h_{(1)}\inprod{F^{-1}}{f_{(2)}g_{(2)}\otimes h_{(2)}}
\inprod{F^{-1}}{f_{(3)}\otimes g_{(3)}}\nonumber\\
&=&f_{(1)}g_{(1)}h_{(1)}\inprod{(\Delta\otimes\id)\left(F^{-1}\right)}{
f_{(2)}\otimes g_{(2)}\otimes h_{(2)}}\inprod{F^{-1}}{f_{(3)}\otimes
g_{(3)}}\nonumber\\
&=&f_{(1)}g_{(1)}h_{(1)}\inprod{(\Delta\otimes\id)\left(F^{-1}\right)
F^{-1}_{12}}{f_{(2)}\otimes g_{(2)}\otimes h_{(2)}}.
\end{eqnarray}
Computing $f*(g*h)$ in a similar fashion replaces the left argument of the
inner product above with $(\id\otimes\Delta)\left(F^{-1}\right)F^{-1}_{
23}$, and if we take the inverse of (\ref{FYBE}), we see the two are equal,
and this proves that $*$ is associative.  The counit condition of $F$
ensures that $1$ is also the $*$-multiplicative identity as well.  We
therefore denote by $\widehat{\H}$ the unital associative algebra with
vector space $\H$ and multiplication $*$.

One consequence of this definition of $*$ is that the Drinfel'd twist of
$\H^*$ is a HA of left actions on $\widehat{\H}$:  Taking $x\in\H^*$ and
$f,g\in\widehat{\H}$,
\begin{eqnarray}
x\cdot(f*g)&=&x\cdot\left(f_{(1)}g_{(1)}\right)\inprod{F^{-1}}{f_{(2)}
\otimes g_{(2)}}\nonumber\\
&=&f_{(1)}g_{(1)}\inprod{x}{f_{(2)}g_{(2)}}\inprod{F^{-1}}{f_{(3)}\otimes
g_{(3)}}\nonumber\\
&=&f_{(1)}g_{(1)}\inprod{\Delta(x)\otimes F^{-1}}{f_{(2)}\otimes g_{(2)}
\otimes f_{(3)}\otimes g_{(3)}}\nonumber\\
&=&f_{(1)}g_{(1)}\inprod{\Delta(x)F^{-1}}{f_{(2)}\otimes g_{(2)}}
\nonumber\\
&=&f_{(1)}g_{(1)}\inprod{F^{-1}\Delta^F(x)}{f_{(2)}\otimes g_{(2)}}
\nonumber\\
&=&f_{(1)}*g_{(1)}\inprod{x_{(1)F}\otimes x_{(2)F}}{f_{(2)}\otimes
g_{(2)}}\nonumber\\
&=&\left(x_{(1)F}\cdot f\right)*\left(x_{(2)F}\cdot g\right).
\end{eqnarray}
So $\H^F$ has a well-defined action on $\widehat{\H}$, and therefore may be
used as a the space of derivations on $\widehat{\H}$.

\bigskip

What if $\H^*$ is quasitriangular?  Then we automatically have an $F$ which
satisfies (\ref{FYBE}),namely, $F=R_{21}^{-1}$.  This follows from the
coproduct properties of the R-matrix, and we recover all the results in
Section \ref{R-leibniz}: We see immediately that the $*$-product given by
plugging $R_{21}$ in for $F^{-1}$ in (\ref{Fstar}) is the same as
(\ref{Rstar}).  The coproduct is also the same, $\Delta^F=\tau\circ
\Delta=\Delta'$.  To compare the antipodes, note that $\sigma^{-1}$ becomes
$m(S\otimes\id)\left(R_{21}\right)$, which is the element, usually denoted
$u$, which generates the square of the antipode in $\H^*$: $uxu^{-1}=S^2(
x)$ \cite{D2}.  This immediately leads to $S^F=S^{-1}=S'$, exactly as
expected, and we see that this choice of $F$ gives $\left(\H^*\right)^F
=\H^{\rm op}$, which, as we proved, is the correct choice for the space of
derivations on $\widehat{\H}$.

\bigskip

There is potentially a wider class of $F$s than there are of R-matrices,
because the one coproduct condition on $F$ (\ref{FYBE}) is less restrictive
than the two coproduct conditions on $R$:
\begin{eqnarray}
(\Delta\otimes\id)(R)=R_{13}R_{12},&&(\id\otimes\Delta)(R)=R_{13}R_{23}.
\label{cop-R}
\end{eqnarray}
This means that, in principle, there may be other associative $*$-products
besides the one defined using the R-matrix.  However, for the
noncommutative string case, the cocommutativity of $\F^*$ is still a strong
enough condition to force $F$ to be identical to the $R$ given in
(\ref{R-matrix}), so in this instance, reformulating the $*$-product on
$\F$ in terms of the Drinfel'd twist does not change the fact that it has
the unique form (\ref{star}).  Therefore, using the Drinfel'd twist gives
exactly the same noncommutativity to the D$p$-brane-open string system as
using an R-matrix does.

\section{Noncommutative String Theory}
\setcounter{equation}{0}

We conclude this work with some brief speculations about how the Drinfel'd
twist may appear in the context of a noncommutative string theory.

\bigskip

In \cite{W}, we conjectured that the dependence on $\theta^{ij}$ in the
effective field theory would appear only through the $*$-product, and that
since this in turn was given in terms of the R-matrix $R$, then there would
be explicit dependence on $R$ in the action when written in terms of the
commutative theory.  However, this is in fact probably not the case in
general, for the following reason: The $*$-product becomes the commutative
product when $\theta^{ij}=0$, which, using the explicit form
(\ref{R-matrix}), corresponds to $R=1\otimes 1$.  When we are dealing with
a cocommutative HA where $\Delta'=\Delta$, this is an admissible
R-matrix, but for the most general case where the HA may {\em not} be
cocommutative, $1\otimes 1$ doesn't work as an R-matrix.

However, the Drinfel'd twist construction is still applicable, because
$F=1\otimes 1$ satisfies (\ref{FYBE}), and just gives the trivial case
$\left(\H^*\right)^F=\H^*$.  Thus, if $\theta$ is some element of a
parameter space, and there exists a continuous map $\theta\mapsto
F(\theta)$ which satisfies (\ref{FYBE}) and $F(0)=1\otimes 1$, we have a
family of spaces $\widehat{\H}(\theta)$ and Drinfel'd twists $\H^F(\theta)$
continuously connected to the undeformed cases $\H$ and $\H^*$,
respectively, with elements of the latter being derivations on elements of
the former.  This deformation does not depend on the cocommutativity, or
lack thereof, of $\H^*$.  Furthermore, since the coproduct of $\H^*$ is
dual to the multiplication on $\H$, this also implies that we do not even
have to start with a commutative $\H$ for this procedure to be valid.

When we look at the specific case of a D$p$-brane/open string system, where
we expect to be able to go continuously from the commutative case with
vanishing $B_{ij}$ to the noncommuting theory, it therefore seems
reasonable to us that the $\theta$-dependence in the effective action when
expressed as an integral over the commutative space will be entirely
through an $F$ and not an $R$, and that the underlying structure is that of
a Drinfel'd twisted HA rather than a quasitriangular one.

\bigskip

There is another reason to favour the Drinfel'd twist: Recall that the
recent work on noncommutative string theory has been done with a {\em
constant} $B_{ij}$.  Since the full theory should be invariant under the
gauge transformation $B_{MN}\mapsto B_{MN}+\pder_{[M}\lambda_{N]}$ for any
$\lambda_M$, where $M,N=0,\ldots,9$, taking $B_{MN}=0$ for $M,N=p+1,
\ldots,9$ and constant for $M,N=0,\ldots,p$ is a gauge fixing condition.
We should therefore expect to find a Ward identity resulting from this
fixing.  To us, it seems very likely that this Ward identity is related to
(\ref{FYBE}).  This explanation is attractive because, if correct, it means
we do not have to impose (\ref{FYBE}) by hand; it comes out naturally from
taking $B_{ij}$ to be constant.  And just as a Ward identity must hold to
have a self-consistent theory, \ie gauge invariance, so must (\ref{FYBE})
hold for consistency, \ie $*$ is associative.  For an R-matrix to be
involved instead of $F$, we would have to come up with some way of arriving
at the {\em two} conditions (\ref{cop-R}), and the single requirement that
the theory have $B$-gauge invariance would presumably not give these.

\bigskip

Thus, the signs point more toward a Drinfel'd twisted rather than a
quasitriangular HA; more specifically, if we think of the `undeformed'
theory as one formulated with the HA $\H$ (commutative or not), and the
`deformed' one as that on the algebra $\widehat{\H}$, then the former
should have explicit $F$-dependence.  This fact may therefore give some
clues as to the explicit form of the undeformed low-energy effective
action, even though it is presumably very complicated (unlike the deformed
version, which may be very nice, \eg super-Yang-Mills \cite{SW}).

\section*{Acknowledgements}

I'd like to thank Florin Panaite for his helpful comments and for pointing
out the references \cite{GZ,BPvO}, and to Robert Oeckl for \cite{O}.

\end{document}